\documentclass[figures,preprint,cite]{epl}

\usepackage{amssymb}
\usepackage{epsfig}

\begin{document}

\title{Non--Ergodic Behaviour of the $k$--Body Embedded\\ Gaussian
Random Ensembles for Bosons}

\author{T.~Asaga \and L.~Benet \and T.~Rupp \and H.~A.~Weidenm\"uller}

\institute{Max--Planck--Institut f\"ur Kernphysik, D--69029
Heidelberg, Germany}


\shorttitle{Non--Ergodic Behaviour in Embedded Gaussian Ensembles}
\shortauthor{T.~Asaga \etal}

\maketitle

\begin{abstract}
  We investigate the shape of the spectrum and the spectral
  fluctuations of the $k$--body Embedded Gaussian Ensemble for Bosons
  in the dense limit, where the number of Bosons $m \to \infty$ while
  both $k$, the rank of the interaction, and $l$, the number of
  single--particle states, are kept fixed. We show that the relative
  fluctuations of the low spectral moments do not vanish in this
  limit, proving that the ensemble is non--ergodic. Numerical
  simulations yield spectra which display a strong tendency towards
  picket--fence type. The wave functions also deviate from canonical
  random--matrix behaviour.
\end{abstract}

\section{Introduction}
\label{int}

Random--matrix theory (RMT) successfully describes the statistical
behaviour of spectra and wave functions of a large variety of systems
such as atoms, molecules, atomic nuclei and quantum
dots~\cite{bro81,guh98}. However, this RMT modeling is not completely
realistic since all many--body systems are effectively governed by
one-- and two--body forces. This fact led to work on the two--body
random ensembles for Fermions~\cite{fre71,boh71,ger72,fre73} and to
the introduction of the $k$--body embedded ensembles by Mon and
French~\cite{mon75}. In the embedded ensembles, many--body states are
constructed by distributing $m$ particles over $l$ degenerate
single--particle levels. The matrix of the $k$--body interaction with
$k \le m$ is taken in this basis. For $k < m$, the $m$--body matrix
elements of the random $k$--body interaction are correlated: The
number of independent random variables is smaller than in RMT. Do
these more realistic embedded ensembles yield the same results as RMT?
Early numerical simulations for interacting Fermion
systems~\cite{fre71,boh71} of rather small matrix dimension have shown
that the spectral fluctuation properties of the embedded ensembles
agree with those of RMT. Similar results were obtained in numerical
simulations for Bosonic systems~\cite{man84,patel00}. Moreover, the
Fermionic ensembles were shown to be ergodic. As far as we know, there
are no results on the spectral ergodicity for Bosons.

Recently, three of the present authors introduced a novel analytical
approach to the Fermionic embedded ensembles in the limit of infinite
matrix dimension ($l \to \infty$)~\cite{ben01a}. The main results of
this approach are: (i)~For $2k > m$, the average spectrum has
semicircle shape, and the spectral fluctuation properties coincide
with those of RMT; (ii)~the spectral density changes shape at or near 
$2k = m$ and becomes Gaussian; (iii)~in the dilute limit ($k \ll m \ll
l$) the spectral fluctuations are completely uncorrelated (Poissonian);
(iv)~the spectral fluctuations change gradually from Wigner--Dyson to
Poisson.

In this Letter, we extended our work to the case of Bosons. As in the
Fermionic case, we consider the limit of infinite matrix
dimension. For Bosons this limit is realized either by letting $l \to
\infty$ (the same limit as for Fermions), or by letting $m \to \infty$
with $k$ and $l$ fixed. This second case, the dense limit, is novel
and has no analogue in the Fermionic case. In the following, we focus
attention exclusively on the dense limit. We prove analytically that
in this limit, the ensemble is not ergodic. Numerical results for the
spectral correlations are obtained by both ensemble unfolding and by
spectral unfolding. In the latter case we find a strong tendency of
the spectra towards picket--fence behaviour. We also show that some
eigenfunctions display Fock--space localization. Details of the
derivations and a complete treatment including the limit $l \to
\infty$, are given in Ref.~\cite{asa01}.

\section{Definitions}
\label{def}

We consider $m$ spinless Bosons in $l$ degenerate single--particle
states with associated creation and annihilation operators
$b_j^{\dagger}$ and $b_j$ where $j = 1,\ldots,l$. Hilbert space is
spanned by the $N={l+m-1 \choose m}$ orthonormal $m$--particle states
$|\mu\rangle$ ($\mu = 1,\ldots,N$), written as $| \mu \rangle = [{\cal
N}(j_1,\ldots,j_m)]^{-1} b_{j_1}^\dag \ldots b_{j_m}^\dag$ with $j_1
\leq j_2 \leq \ldots \leq j_m$ and ${\cal N}(j_1,\ldots,j_m)$ a
normalization constant. The states $|\mu\rangle$ are equivalently
characterized by a sequence of occupation numbers $(n_1,\ldots,n_l)$
of the $l$ single--particle states with $b_{j_1}^\dag \ldots
b_{j_m}^\dag = (b_1^\dag)^{n_1} \ldots (b_l^\dag)^{n_l}$. Setting
${\cal N}(j_1,\ldots,j_m) = \sqrt{n_1! \ldots n_l!}$ normalizes all
$m$--particle states $|\mu\rangle$ to 1.

The bosonic $m$--particle states are coupled through a random
$k$--body interaction $V_k(\beta)$ with $k \leq m$, given by
\begin{equation}
\label{eq2.1}
V_k(\beta) = \sum_{{1 \leq j_1 \leq j_2 \leq \ldots \leq j_k \leq l}
  \atop {1 \leq i_1 \leq i_2 \leq \ldots \leq i_k \leq l}}
  v_{j_1,\ldots,j_k; i_1,\ldots,i_k} \,
  {b_{j_1}^{\dagger} \ldots b_{j_k}^{\dagger} b_{i_k} \ldots b_{i_1}
  \over {\cal N}(j_1 \ldots j_k) {\cal N}(i_1 \ldots i_k) } \ .
\end{equation}
We refer to $k$ as to the rank of the interaction. As in the canonical
case, the labels $\beta = 1$ and $\beta = 2$ denote the orthogonal and
the unitary ensemble, respectively. The matrix element
$v_{j_1,\ldots,j_k; i_1,\ldots,i_k}$ of the $k$--body interaction
taken between the single--particle states $j_1,\ldots, j_k$ and
$i_1,\ldots, i_k$ is totally symmetric with respect to $j_1,\ldots,
j_k$ and $i_1,\ldots, i_k$. The elements differing in the sequence of
indices $\{j_1 \ldots j_k; i_1 \ldots i_k \}$ (except for permutations
of $\{j_1 \ldots j_k\}$ and of $\{i_1 \ldots i_k \}$ and for
symmetries specified by $\beta$) are uncorrelated
Gaussian--distributed random variables with zero mean and a common
second moment $v_0^2$. Without loss of generality we put $v_0^2=1$ in
the sequel. The normalization coefficients ${\cal N}(j_1,\ldots,j_k)$
and ${\cal N}(i_1,\ldots,i_k)$ in Eq.~(\ref{eq2.1}) guarantee that for
$k = m$, the embedded ensembles are identical to the canonical
ensembles of random--matrix theory. This defines the Bosonic $k$--body
embedded Gaussian orthogonal (unitary) ensemble of random matrices,
respectively, in short BEGOE($k$) and BEGUE($k$). The number of
independent random variables is given by $K_{\beta} = \beta {l+k-1 
 \choose k}[{l+k-1 \choose k }+\delta_{\beta 1}]/2$.

\section{The Second Moment}
\label{bas}

By virtue of the randomness of $V_k(\beta)$, the elements of the
matrix $\langle \nu | V_k(\beta) | \mu \rangle$ are random variables
with a Gaussian probability distribution and zero mean value. The
spectral properties are completely determined by the second moment
\begin{eqnarray}
\label{eq3.1}
B^{(k)}_{\mu \nu, \rho \sigma} (\beta)&=& \overline{\langle \mu |
  V_k(\beta) | \sigma \rangle \langle \rho | V_k(\beta) | \nu
  \rangle}\nonumber \\ &=& \sum_{\alpha (k), \gamma(k)} \langle \mu |
  {\cal B}_{\alpha (k)}^{\dagger} {\cal B}_{\gamma(k)} | \sigma
  \rangle \Bigl[ \langle \rho | {\cal B}_{\gamma(k)}^{\dagger} {\cal
  B}_{\alpha (k)} | \nu \rangle + \delta_{\beta 1} \langle \nu | {\cal
  B}_{\gamma(k)}^{\dagger} {\cal B}_{\alpha (k)} | \rho \rangle \Bigr]
  \ .
\end{eqnarray}
The overbar denotes the average over the ensemble. We have simplified
the notation by introducing the operators ${\cal B}_{\alpha
  (r)}^\dagger = [{\cal N}(j_1,\ldots,j_r)]^{-1} b_{j_1}^\dagger
\ldots b_{j_r}^\dagger $ and the adjoints ${\cal B}_{\alpha(r)}$. The
index $\alpha(r)$ is a short--hand notation for the rank $r$ and for
the sequence of indices $\{j_1 \ldots j_r\}$. The matrix
$B^{(k)}(\beta)$ is Hermitean in the pairs of indices $(\mu,\nu)$ and
$(\rho,\sigma)$. It is easy to prove the ``duality'' relation
$B^{(k)}_{\mu \nu,\rho\sigma}(2) = B^{(m-k)}_{\mu\sigma,\rho\nu}(2)$
which for $\beta = 2$ connects the second moments of the $k$--body and
the $(m-k)$--body interactions. For $k = m$ and $\beta = 2$, the
right--hand side of Eq.~(\ref{eq3.1}) reduces to $\delta_{\mu \nu}
\delta_{\rho \sigma}$, and correspondingly for $\beta = 1$. This shows
that for $k = m$, BEGUE($k$) and BEGOE($k$) reduce to the GUE and GOE,
respectively.

In the Fermionic case, results for the shape of the spectrum and for
the spectral fluctuations were obtained with the help of the
eigenvector decomposition of the second moment. A similar
decomposition exists in the case of Bosons and is derived in
Ref.~\cite{asa01}. In the following, we only state the results of this
derivation. We solve the eigenvalue equation $\sum_{\rho\sigma}
B^{(k)}_{\mu \nu, \rho \sigma}(2) C^{(sa)}_{\sigma \rho} =
\Lambda^{(s)}(k) C^{(sa)}_{\mu \nu}$, with eigenvectors $C^{(sa)}$ and
with eigenvalues $\Lambda^{(s)}(k)$ given by
\begin{equation}
\label{eq4.4}
\Lambda^{(s)}(k) = {m-s \choose k} {l+m+s-1 \choose k} \ .
\end{equation}
Here, $s=0,\ldots,m$ while $a$ labels the degenerate eigenvectors. For
the degree of degeneracy $D^{(s)}$ of the eigenvalues
$\Lambda^{(s)}(k)$ we find $D^{(0)}=1$ and $D^{(s)} = {l+s-1 \choose
s}^2 - {l+s-2 \choose s-1}^2$ for $s \geq 1$. It follows that
$\sum_{s=0}^m D^{(s)} = N^2$ showing that the eigenvectors form a
complete set. We choose Hermitean linear combinations of the
degenerate eigenvectors which obey the orthonormality condition
$\sum_{\mu\nu} C^{(sa)}_{\mu\nu} C^{(tb)}_{\nu\mu} = N \delta_{st}
\delta_{ab}$. These results can be extended to $\beta=1$. Hence, the
matrix $B^{(k)}(\beta)$ can be expanded as
\begin{equation}
\label{eq4.6}
B^{(k)}_{\mu\nu,\rho\sigma}(\beta) = \frac{1}{N} \sum_{s=0}^m
\Lambda^{(s)}(k) \sum_a [C^{(sa)}_{\mu\nu} C^{(sa)}_{\rho\sigma} +
\delta_{\beta 1} C^{(sa)}_{\mu\rho} C^{(sa)}_{\nu\sigma}]\,.
\end{equation}

\section{Low Moments of $V_k$}
\label{low}

Using the eigenvector decomposition of $B^{(k)}(\beta)$, duality and the
orthonormality of the $C^{(sa)}$'s, explicit expressions for the low moments
of $V_k(\beta)$ are obtained. From these we calculate three ratios that yield
information on the shape of the spectral density. The ratio $S$ measures
the fluctuations of the center of the spectrum in units of the average width
of the spectrum. The ratio $R$ measures the relative fluctuations of the width
of the spectrum. The ratio $Q$ is related to the kurtosis $\kappa = Q
+ 2$ and marks the difference between the semicircular ($Q=0$) and the
Gaussian shape ($Q=1$). The definitions of $S$, $R$, and $Q$ are given in
Ref.~\cite{ben01a}.

We are particularly interested in the behaviour of the three ratios in
the dense limit, i.e. the case $N \to \infty$ attained by letting $m
\to \infty$ while keeping both $l$ and $k$ fixed. We obtain
\begin{eqnarray}
\label{eq5.6}
\lim_{m\to \infty}S(k,m,l)&=&{(1+\delta_{\beta 1}){2k \choose k}
  {l+k-1 \choose k}^{-1} \over {2k \choose k} + \delta_{\beta 1} 
  \sum_{s=0}^k {2k \choose k+s} {l+k+s-1 \choose k+s}^{-1} 
  d^{(s)}}\ , \\
\label{eq5.7}
\lim_{m\to \infty} R(k,m,l) &=& { 2\sum_{s=0}^k [{2k \choose k+s}
  {l+k+s-1\choose k+s}^{-1}]^2 \bigl[D^{(s)} +\delta_{\beta 1}
  (D^{(s)} + 2d^{(s)}) \bigr] \over \Bigl[{2k \choose
  k} +\delta_{\beta 1} \sum_{s=0}^k {2k \choose k+s} {l+k+s-1 \choose
  k+s}^{-1} d^{(s)}\Bigr]^2} \ , \\
\label{eq5.8}
\lim_{m\to \infty} Q(k,m,l) &=& \sum_{s=0}^k {2k \choose k}^{-1} 
  {2k \choose k+s} {l+k+s-1 \choose k+s}^{-1} D^{(s)} 
  = 1 \, ,
\end{eqnarray}
with $d^{(s)} = {l+s-2 \choose s}$. 
Eq.~(\ref{eq5.8}) applies only in the unitary case ($\beta=2$) and
implies that in the dense limit, the average spectrum has Gaussian
shape. We have not been able to extend this analytical result to the
orthogonal case $\beta=1$. On physical grounds, however, an equation
analogous to Eq.~(\ref{eq5.8}) is expected to be valid also for the
BEGOE($k$). We conclude that the shape of the average spectrum has
Gaussian form. This is in keeping with the results of Ref.~\cite{kot80}.

A more important and surprising result lies in the fact that the
right--hand sides of Eqs.~(\ref{eq5.6}) and~(\ref{eq5.7}) do not
vanish: The fluctuations of the centroids and of the variances of
individual spectra do not vanish asymptotically. This feature differs
from the behaviour both of canonical RMT and of the embedded Fermionic
ensembles. We are led to the important conclusion that the Bosonic
ensembles are not ergodic in the dense limit $m \to \infty$ with $k$
and $l$ fixed. Non--ergodicity appears to be a consequence of the
fact that the number of independent random variables $K_\beta$ in the
ensemble does not grow with $m$, but stays finite for fixed $k$ and
$l$.

\section{Numerical results}
\label{num}

For lack of analytical techniques, we use numerical simulations to
obtain infromation on the spectral fluctuation properties of
BEGOE($k$) in the dense limit. With $l=2$ and $k=2$, the dimension of
Hilbert space is given by $N = m + 1$, and $m \gg l$ is easily
attainable numerically. The number of independent random variables is
$K_1=6$. The many--particle states $|\mu_n \rangle$ are written as
$(m-n, n)$ where $m-n$ and $n$ indicate the number of Bosons occupying
the first and the second single--particle state, respectively. In the
basis $\{|\mu_0\rangle, |\mu_1\rangle, \ldots, |\mu_m \rangle\}$, the
Hamiltonian matrix attains band structure, with non--zero matrix
elements on the main diagonal and on the $k$ closest diagonals.

\begin{figure}
\twofigures[width=4.5cm,angle=90]{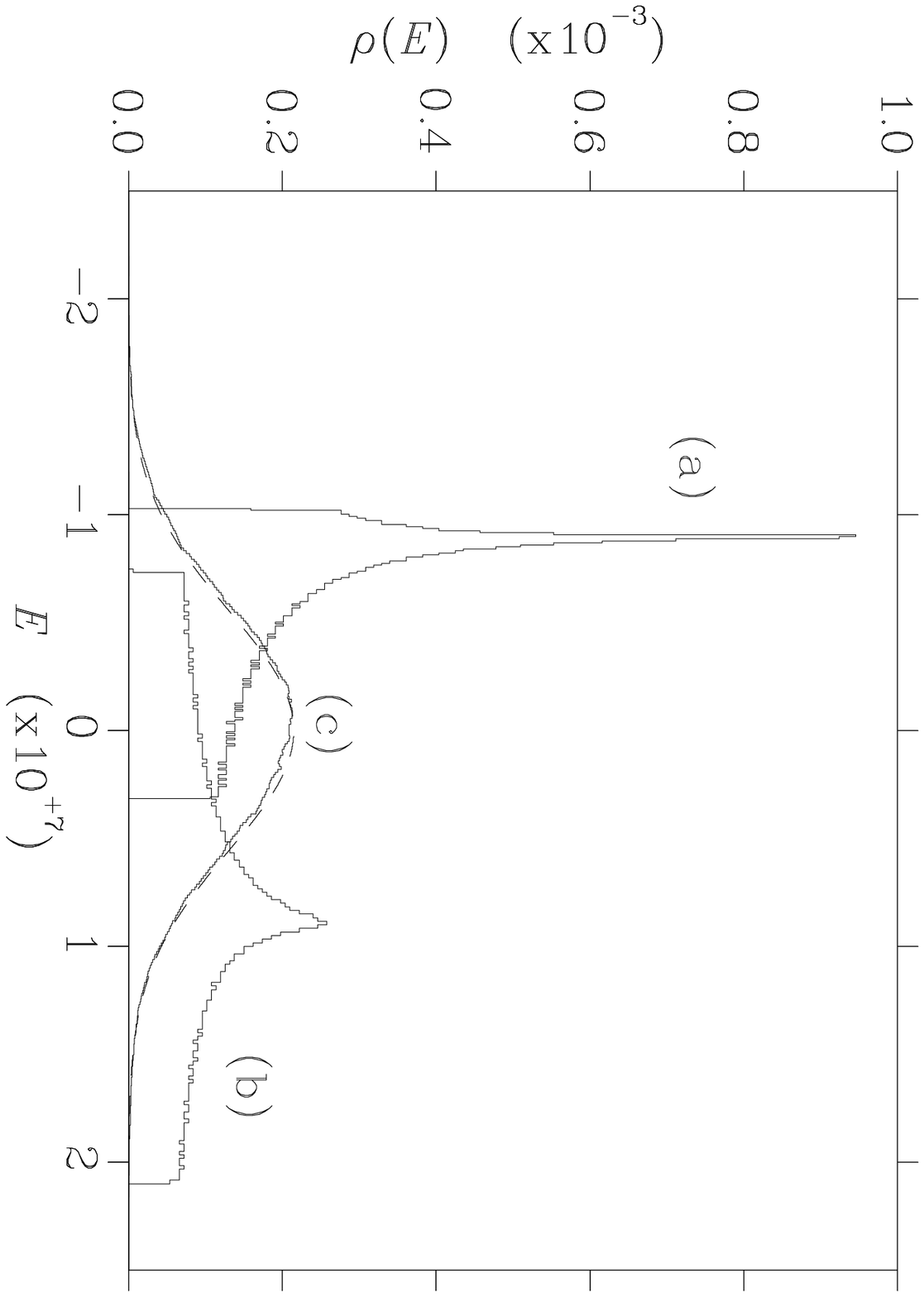}{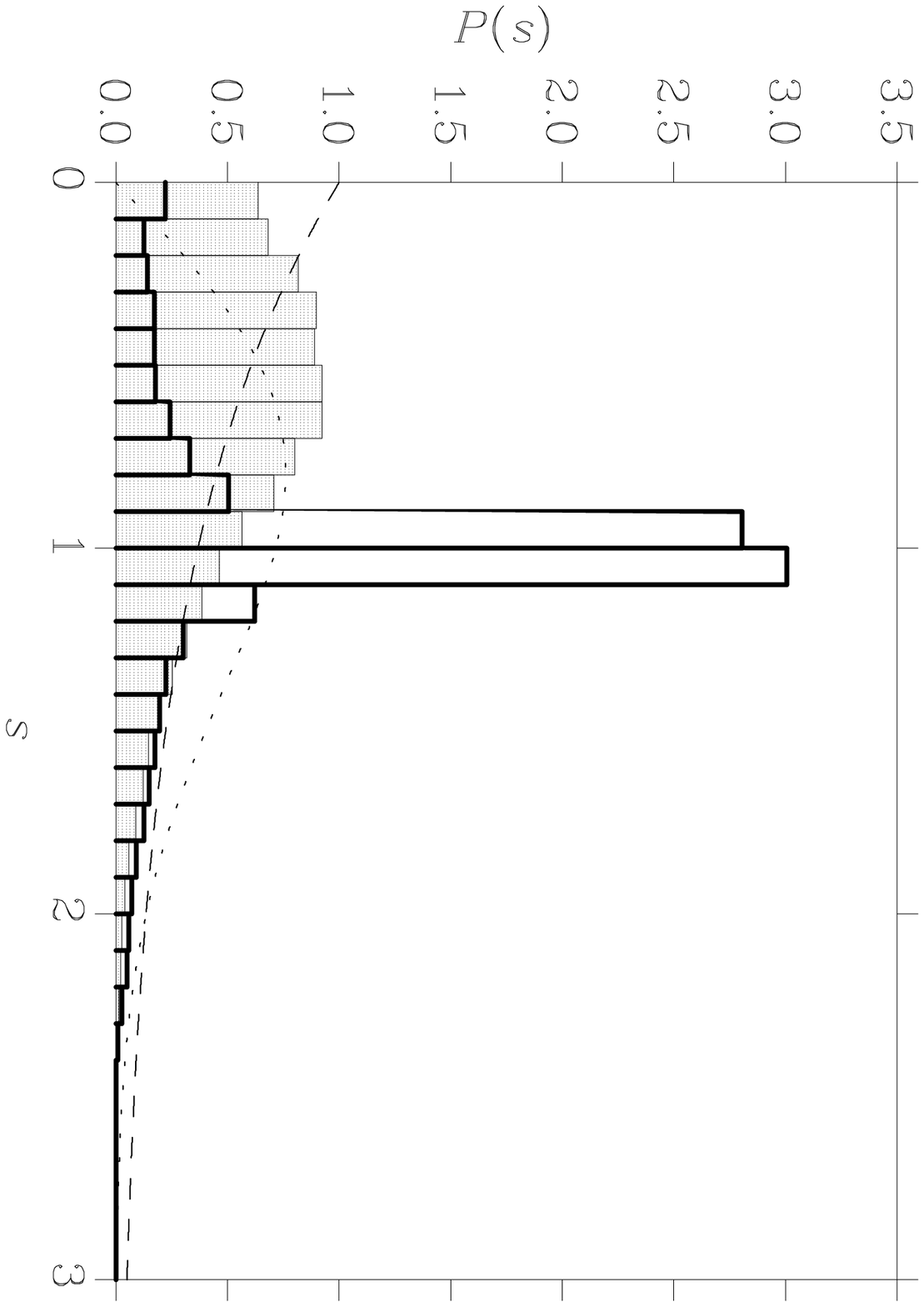}
\caption{Spectral density of the BEGOE($2$) for $l = 2$ and $m =
  3000$, normalized to the dimension $N = 3001$ of Hilbert space. The
  curves (a)~and (b)~show the level densities of two members of the
  ensemble; curve (c)~shows the ensemble--averaged spectral
  density. The dashed curve shows the theoretically predicted
  Gaussian shape of the average spectrum.}
\label{fig1}
\caption{Nearest--neighbour spacing distribution $P(s)$ obtained by
  ensemble unfolding (filled histogram) and spectral unfolding (empty
  histogram). The dotted curve gives the Wigner surmise and the dashed
  one the Poisson distribution.}
\label{fig2}
\end{figure}

Fig.~\ref{fig1} shows the level densities of two members of the
ensemble and the ensemble--averaged density obtained from $1512$
spectra ($m=3000$). We have also plotted a Gaussian density whose
width is given by the theoretical prediction~\cite{asa01} and which
agrees with the shape of the ensemble--averaged density. The striking
differences between each of the two spectral densities and between
these and the ensemble--averaged spectral density illustrate the
non--ergodic character of BEGOE(2) for $m \gg l$. The Gaussian
form of the ensemble--averaged spectrum arises as a consequence of the
Central Limit Theorem. It has no bearing on individual spectral shapes.

The evaluation of measures of spectral fluctuations requires the
spectra to be unfolded. Unfolding can be done either individually for
each spectrum (spectral unfolding) or by a single transformation which
is applied to all spectra of the ensemble (ensemble unfolding). We
employed both types of unfolding in our statistical analysis. For
spectral unfolding, a polynomial was fitted to the staircase function
of each realization of a spectrum. Starting from the value unity, we
changed the degree of the polynomial for each realization until the
first minimum of the associated $\chi^2$ distribution was reached. The
maximum degree considered was 20. The first minimum was typically
found around degree 11. Ensemble unfolding was carried out by first
averaging the staircase functions over the ensemble. The averaged
staircase function was fitted by a polynomial of degree 11. We have
analyzed $1512$ realizations for $m=3000$. After fitting the
polynomials by including all levels of each spectrum for both spectral
and ensemble unfolding, we only considered $60\%$ of all levels, i.e.,
$1800$ levels for each realization located in the centre part of each
spectrum.

\begin{figure}
\twofigures[width=4.5cm,angle=90]{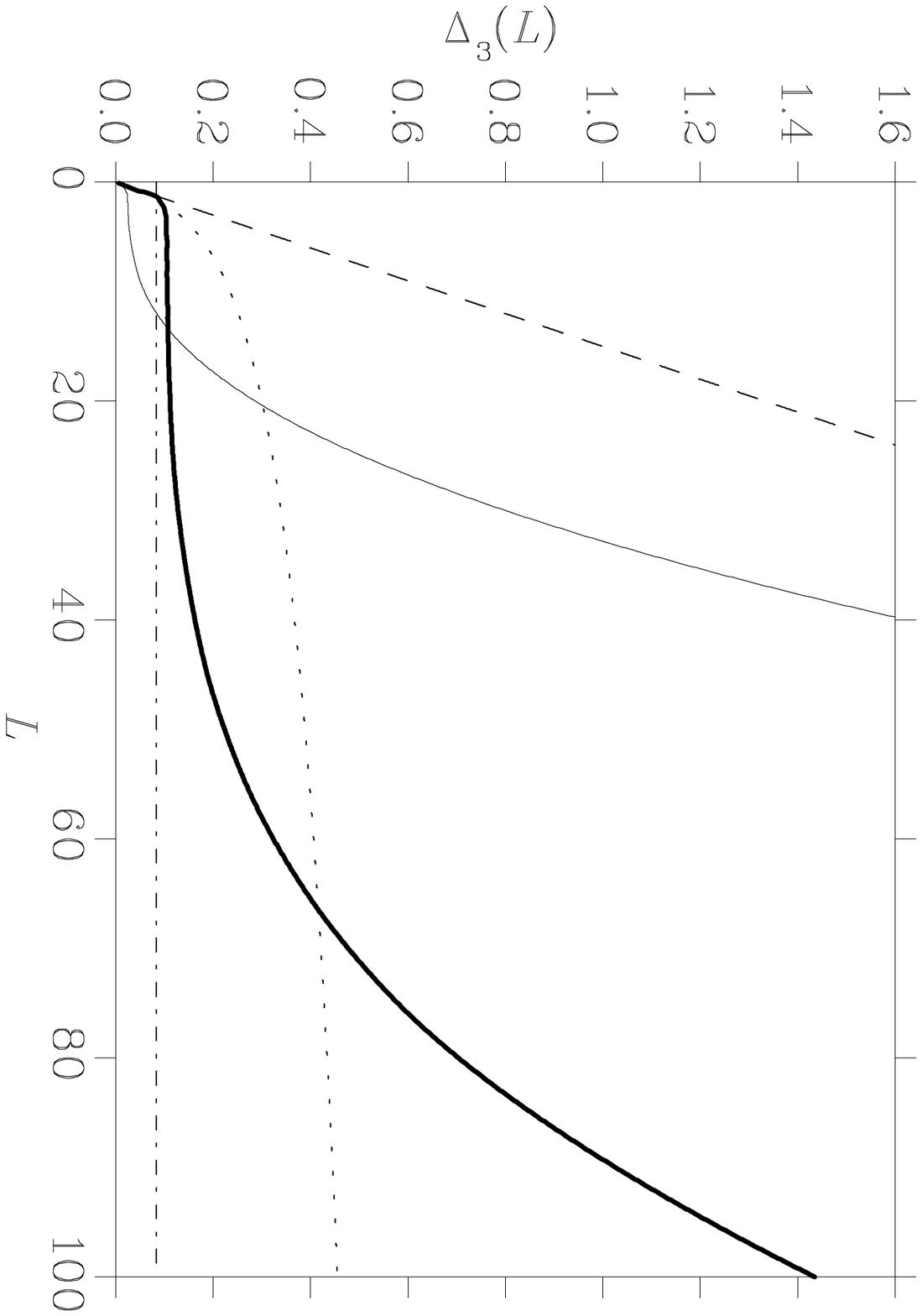}{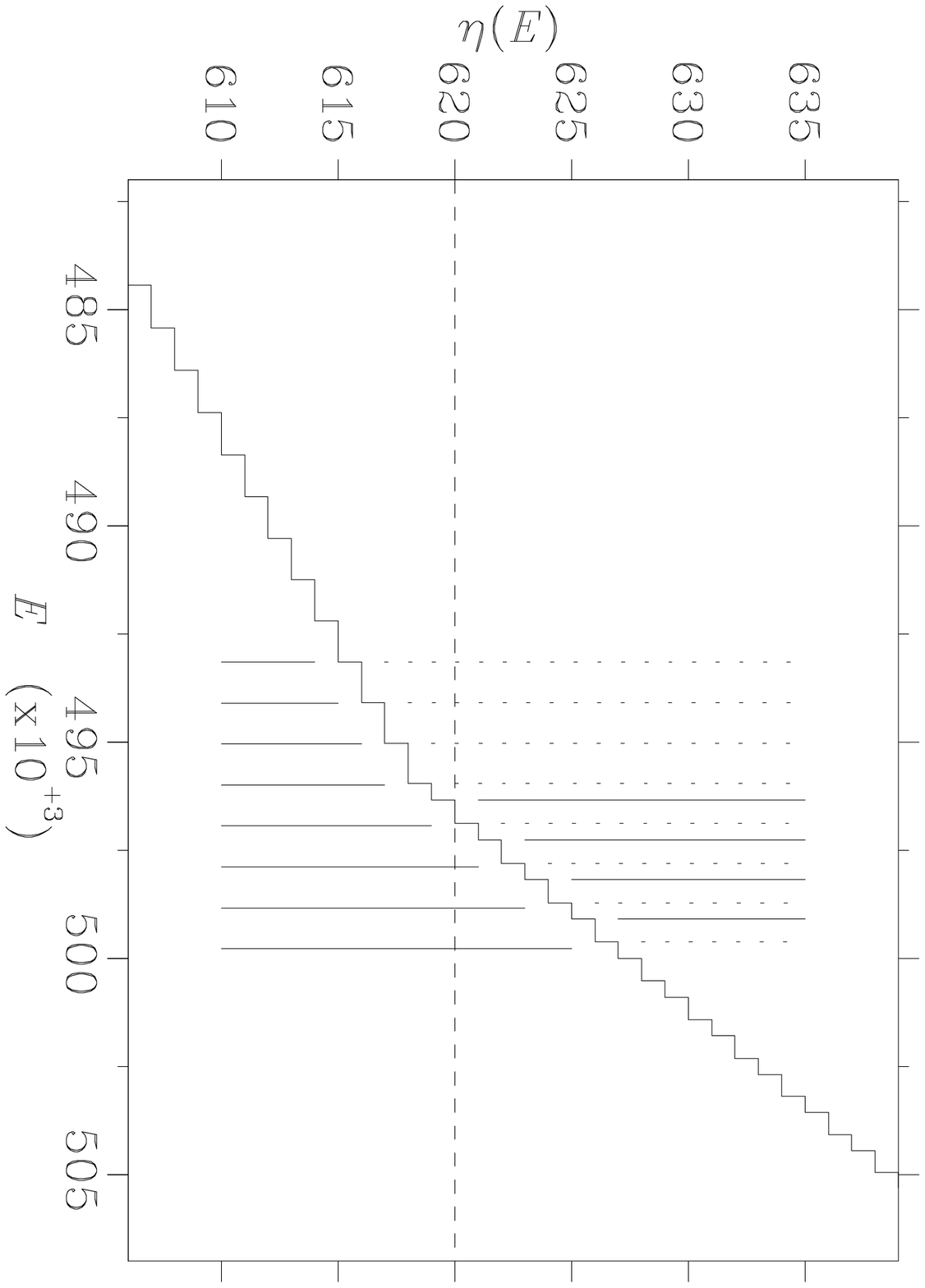}
\caption{ $\Delta_3$--statistics measured at the centers of the
  spectra after ensemble unfolding (thin solid line) and spectral
  unfolding (bold line). For comparison we have plotted the results
  for the GOE (dotted curve), for a Poissonian spectrum (dashed line),
  and for a picket--fence spectrum (dotted--dashed line).}
\label{fig3}
\caption{ Detail of the staircase function $\eta(E)$ for a typical
  member of BEGOE(2) with $m = 1000$ and $l = 2$. The vertical lines
  help to show how two (almost) equidistant spectra overlap around the
  level $620$. We note the change in the density of states after this
  level.}
\label{fig4}
\end{figure}

In Figs.~\ref{fig2} and \ref{fig3} we show the nearest--neighbour
spacing distribution $P(s)$ and the $\Delta_3$--statistics obtained
for $m = 3000$ by ensemble unfolding and by spectral unfolding. In
both cases the distribution $P(s)$ corresponds neither to the Wigner
surmise nor to a Poisson distribution. In the case of ensemble
unfolding the level repulsion characteristic of the GOE is clearly
lost. For spectral unfolding the distribution $P(s)$ is dominated by a
prominent peak centered at $s=1$ suggesting that individual spectra
have an almost constant level spacing. The $\Delta_3$--statistics
obtained by ensemble unfolding deviates rapidly from GOE behaviour
and increases almost linearly. In the case of spectral unfolding,
$\Delta_3(L)$ is almost constant up to $L \sim 20$. This result again
suggests a tendency of individual spectra towards a
picket--fence--like behaviour. Beyond this range, $\Delta_3(L)$ grows
albeit less rapidly than for ensemble unfolding.

We turn to the detailed structure of individual
spectra. Fig.~\ref{fig4} shows the staircase function $\eta(E)$ for a
typical member of the ensemble ($k = 2$, $l = 2$, $m = 1000$). The
spectrum is dominated by levels with almost constant spacing. The
staircase function typically displays one or more points where an
abrupt change in the density of states takes place. In
Fig.~\ref{fig4}, the spectrum has (almost) constant spacing up to the
level $619$. From level $620$ on, the spacing between neighbouring
levels is no longer constant. However, the spacing between
next--to--nearest--neighbours does have this property and is almost
the same as the nearest--neighbour spacing before level $620$. This is
illustrated by the vertical lines plotted above and below the
staircase function. These observations imply non--stationary
properties of the spectra. In addition, our result suggests that the
spectrum of an individual member of the ensemble consists of pieces of
overlapping segments of spectra with almost constant level spacings.
The kink at level $620$ in Fig.~\ref{fig4} marks the left edge of the
overlap region.

\begin{figure}
\onefigure[width=10cm]{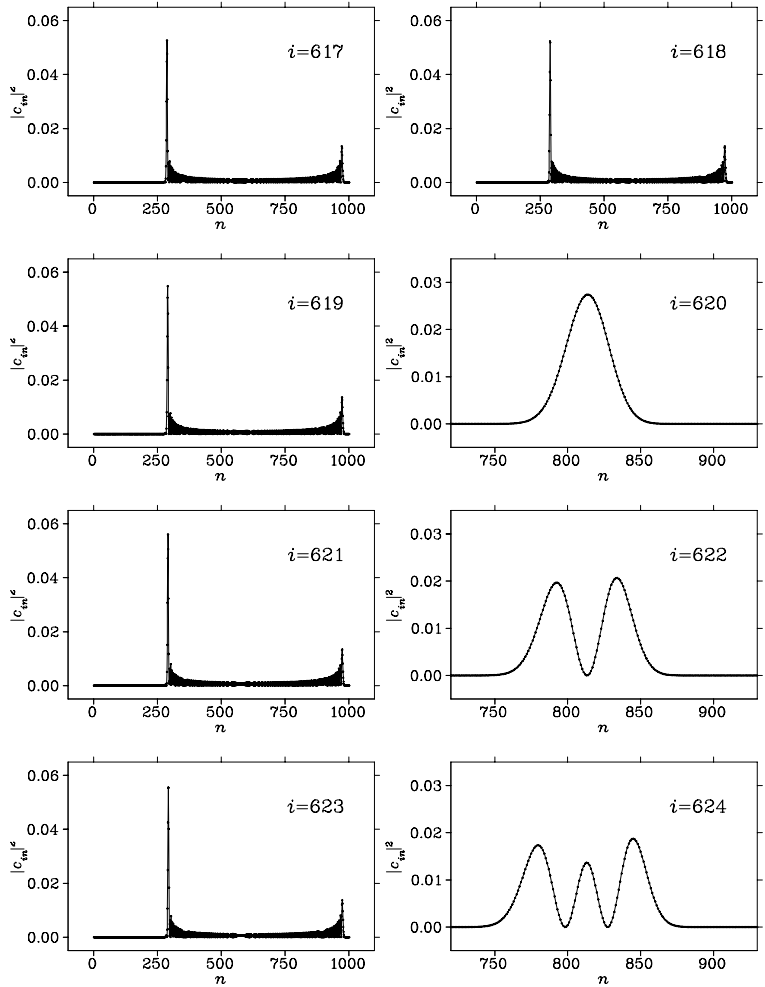}
\caption{ Probabilities $|c_{in}|^2$ for eigenvectors belonging to
  eigenvalues in the vicinity of the kink shown in
  Fig.~\ref{fig4}. The two overlapping segments of nearly equidistant
  levels are easily distinguished by the structure of the
  eigenfunctions.}
\label{fig5}
\end{figure}

The overlap of two segments of almost equidistant levels also
influences the structure of the eigenfunctions. We write the
eigenvectors $| i \rangle$ as linear combinations of the ordered
many--body basis states $| \mu_n \rangle$, $| i \rangle = \sum_{n=0}^m
c_{in} | \mu_n \rangle$. In Fig.~\ref{fig5} we plot the probabilities
$|c_{in}|^2$ of eigenvectors $| i \rangle$ belonging to eigenvalues in
the vicinity of the kink shown in Fig.~\ref{fig4}. Eigenvectors up to
$i = 619$ behave similarly and are somewhat extended, although the
distribution of the intensities clearly deviates from the
Porter--Thomas distribution expected from RMT. This behaviour changes
abruptly at $i = 620$. We emphasize the difference in the scales used
in Fig.~\ref{fig5} for $i = 619$ and for the even $i$ values starting
with $i = 620$. The eigenvectors with $i = 621, 623$ display the same
behaviour as the eigenvectors up to $i=619$ and, thus, correspond to
the first segment of the two equidistant spectra. The eigenvectors
with $i =$ 620, 622, 624 are much more localized in Fock space. They
differ in the number of intensity oscillations. As $i$ increases, so
does this number, and the levels on this second segment become more
and more delocalized. At some point the spread of the eigenfunctions
in the second segment is indistinguishable from that in the first one.

\section{Conclusions}
\label{con}

We have studied the spectral behaviour of the Bosonic embedded
ensembles in the dense limit attained by letting $m \to \infty$ with
$l$ and $k$ fixed. We have shown that in this limit, both the
BEGOE($k$) and the BEGUE($k$) are non--ergodic. To the best of our
knowledge, this is the first case of a random--matrix model for which
such non--ergodic behaviour has been established in the limit of
infinite matrix dimension. Moreover, we have shown that the spectral
fluctuations deviate strongly from RMT results. This result disagrees
with the conclusions based on numerical simulations of
Refs.~\cite{man84,patel00}. We ascribe this disagreement to the fact
that the ratios $m/l$ considered in Refs.~\cite{man84,patel00} were
too small to produce deviations from Wigner--Dyson statistics. Our
numerical simulations provide evidence for both a non--stationary and
a picket--fence--type of behaviour in both short-- and long--range
correlations of the spectra. More precisely, individual spectra seem
generically to consist of overlapping segments of spectra of
picket--fence type. In the overlap region, the eigenfunctions of the
levels in the two segments differ markedly, the eigenfunctions in the
second segment displaying strong localization in Fock space. We have
not yet attained an analytical understanding of these properties which
we attribute to the small number of independent random variables
$K_\beta$ (with $K_\beta$ independent of $m$), and to the specific
structure of the Hamiltonian matrix of the Bosonic ensembles. In
particular, we do not know how the spectra change as $l$ increases.

\acknowledgments We are grateful to O. Bohigas and T.H. Seligman 
for stimulating discussions and useful suggestions. T.A. acknowledges
support from the Japan Society for the Promotion of Science.

\end{document}